\documentclass [12pt] {article}
\usepackage{graphicx}
\begin {document}
\baselineskip=24pt
\parskip 10pt plus 1pt
\vskip 2pc \centerline {\large{\bf Influence of an external
    magnetic field on the decoherence of a }}
\vskip 0.2pc \centerline{\large{\bf central spin coupled to an
antiferromagnetic environment }}

{Xiao-Zhong Yuan}$^{1,2}$, Hsi-Sheng Goan$^1$* and Ka-Di
Zhu$^2$

\textit{$^1$Department of Physics and Center for Theoretical
Sciences, National Taiwan University, and 
National Center for Theoretical Sciences, Taipei 10617, Taiwan}

\textit{$^2$ Department of Physics, Shanghai Jiao Tong University,
Shanghai 200240, China}

\begin {center}
\begin {minipage} {5.0 in}
\textbf{Abstract}. Using the spin wave approximation, we study the
decoherence dynamics of a central spin coupled to an
antiferromagnetic environment under the application of an external
global magnetic field. The external magnetic field affects the
decoherence process through its effect on the antiferromagnetic
environment. It is shown explicitly that the decoherence factor
which displays a Gaussian decay with time depends on the strength
of the external magnetic field and the crystal anisotropy field in
the antiferromagnetic environment.
When the values of the external magnetic field is
increased to the critical field point at which the spin-flop
transition (a first-order quantum phase transition)
happens in the antiferromagnetic environment, the decoherence
of the central spin reaches its highest point. 
This result is consistent with several recent
quantum phase transition witness studies. 
The influences of the environmental
temperature on the decoherence behavior of the central spin are
also investigated.
 \vskip 2pc
PACS number(s): 03.65.Yz, 03.67.-a, 75.30.Ds
\end{minipage}
\end{center}
  *e-mail: \textit{goan@phys.ntu.edu.tw }

 \vfill \eject \parindent 1.5pc \vskip 0.5pc
\centerline{\large{\bf I. Introduction}}

Maintaining sufficient quantum coherence and quantum superposition
properties is one of the most important requirements for
applications in quantum information processing, such as quantum
computing, quantum teleportation [1,2], quantum cryptography [3],
quantum dense coding [4], and telecloning [5]. Decoherence is a
process by which a quantum superposition state decays into a
classical, statistical mixture of state. It arises from the
unavoidable interaction and thus entanglement between quantum
systems and the environments. When the environmental degrees of
freedom are traced out or averaged over, the quantum states of the
quantum systems are no longer pure and the quantum systems are
said to have decohered.

In recent years, the decoherence behavior of a small quantum system,
particularly a spin system, interacting with a large
environment has attracted extensive attentions [6-17]. Different
environments result in different decoherence features [18].
In some cases of the problems of a central spin coupled to an environment,
a spin environment may be more appropriate to represent
the actual localized background spins or magnetic defects than the
delocalized oscillaor bath that is usually used.
The spin environment could be lattice nuclear spins,
Ising spin bath, spin-star environment or the device substrates
doped with spin impurities.  In Ref. \cite{eAbe}, the decoherence behavior
of a localized electron spin in the nuclear spin environment is
investigated experimentally. The influence of
an external magnetic field is also considered. The decoherence
factor is shown to be a Gaussian type. By attributing the electron
spin decoherence to quantum fluctuations of lattice nuclear spins,
Ref. [20] attempts to explain the experimental results. There, the
significant coupling between the electron spin and nuclear spins
is assumed to be the Ising-type interaction and the coupling among
nuclear spins is, however, taken to be the dipole-dipole
interaction. 
Recently, the dynamics of the reduced density matrix
for central spin systems in a spin bath described by the
transverse Ising model has been analyzed using a perturbative
expansion method [16] or a mean-field approximation [17,21]. The
interaction between the central spin system and the spin bath for
these case was assumed to be of a Ising type. 
It has also been
reported recently that the reduced dynamics and for one- and
two-spin-qubit systems in a spin-star environment \cite{Bose04}
has been analyzed [23-26]. There, the interaction between the
system and environment was assumed to be of a Heisenberg $XY$
interaction.

Antiferromagnetic materials have been reported recently to have
applications in the area of quantum information processing
[27-32]. It was shown that quantum computing is possible with a
wide variety of clusters assemble from antiferromagnetically
coupled spins [28]. An antiferromagnetic molecular ring was also
suggested [32] to be a suitable candidate for the qubit
implementation. The decoherence of the cluster-spin degrees of
freedom in the antiferromagnetic ring is expected to arise mainly
from the hyperfine coupling with the nuclear spins.
Reference [30] investigated the electron spin decoherence rate of the
quantum tunneling of the N\'{e}el vector in a ring type
antiferromagnetic molecular magnet through the study of a nuclear
spin coupled to an electron spin in the ring. In this paper, we
study the opposite case. We investigate the decoherence behavior
of a central spin coupled to an antiferromagnetic environment.
This decoherence problem of a central spin in an antiferromagnetic
environment has been investigated in Ref.~\cite{33}. A similar problem of
a spin-1/2 impurity embedded in an antiferromagnetic environment has
been studied in the context of quantum frustration of decoherence 
in Ref.~\cite{Novais05}. There the impurity spin is coupled in real space 
locally to just one spin of the antiferromagnetic environment, in
contrast to the central spin 
model where the central spin is coupled isotropically to all the spins
of the environment. In Ref.~\cite{Kohler06}, 
a large-spin impurity in a ferromagnetic bath is
investigated in the context of quasi-classical partial frustration of
dissipation.  

In the present paper, we consider
the case of how an external magnetic field will affect the
decoherence behavior of a central spin in an antiferromagnetic
environment.  One of the
interesting phenomena in antiferromagnetic materials under an applied
magnetic field is the magnetic-field-induced spin-flop transition.
When the applied magnetic field is increased to the critical field point,
the antiferromagnetic polarization flips into the direction
perpendicular to the field. This is called the spin-flop transition.
The phenomena of the spin-flop transition have been observed
experimentally \cite{36,37}.
It is thus particularly interesting to investigate how the central spin
coherence changes as a function of a globally applied external field
in an antiferromagnetic environment,
especially when the magnetic field strength is near the critical field
of the flip-flop transition.
Indeed, the study of the relationship  
between the pure-mixed state transition of a central spin
and the quantum phase transition (QPT) of a bath to which the central
spin is coupled has attracted much attention 
recently \cite{Quan06,Cucchietti07,Rossini07}.  
A QPT is driven only by quantum fluctuations
(e.g., happens at zero temperature where the thermal fluctuations vanish)
and can be triggered by a change of some of the coupling constants
(e.g., external fields) defining the system's Hamitonian.  
In Refs.~\cite{Quan06,Cucchietti07,Rossini07} 
the concept of Loschmidt echo (or equivalently the
decoherence factor) has been used to investigate quantum
criticality in the XY and anisotropic Heisenberg spin chain models. 
While the quantum state fidelity approach has also been put
forward in Refs.~\cite{Zanardi06,Zanardi07} 
to study QPTs in the Dick, BCS-like and XY models. 
Basically, the signature of QPT (in the case of finite spin number)
is the dramatic decay of the asymptotic value of the Loschmidt echo 
or the quantum state fidelity at the critical
point. That is, the closer the
bath to the QPT, the smaller the
asymptotic value of the Loschmidt echo or the
decoherence factor of the central spin. 
The extension of the investigations of quantum criticality 
to finite temperatures has been reported in Ref.~\cite{Zanardi07}.  
We will show in this paper that a similar behavior of the decoherence factor
of the central spin also appears in our antiferromagnetic bath model when
the external magnetic field strength is tuned to near the critical field point 
of the flip-flop transition, a first order QPT.

It is known that the values of the
critical magnetic field of the spin-flop transition can be
obtained using the spin wave theory \cite{43,44,45}. We thus
apply, in this paper,
the spin wave approximation to deal with the antiferromagnetic
environment in a globally applied magnetic field
at low temperatures and low energy excitations.
We obtain explicitly an analytic
expression of the decoherence factor for the central spin. It is
found that the external magnetic field affects the decoherence
process through its effect on the antiferromagnetic environment.
Our results explicitly show that the decoherence factor which
displays a Gaussian decay with time depends on the strength of the external
magnetic field and the crystal anisotropy field in the
environment. One of the important results of our present paper is
that when the magnetic field is increased to the critical field point
at which the spin-flop transition happens, the decoherence of
the central spin reaches its
highest point, consistent with the QPT witness studies 
in Ref.~\cite{Quan06,Cucchietti07,Rossini07,Zanardi07}. 
We also investigate the influence of the
environmental temperature on the decoherence of the central spin.

The paper is organized as follows. In Sec.~II the model
Hamiltonian is introduced and the spin wave approximation is applied
to map the spin operators of the antiferromagnetic
environment to bosonic operators.
In Sec.~III, by tracing over the environmental degrees of freedom,
we calculate the time evolution of the off-diagonal density matrix
elements of the central spin and obtain explicitly the
decoherence time of the central spin. In Sec.~IV, the influence
of the environmental temperature, the environmental structure, and
the external magnetic field on the decoherence behavior of the
central spin are discussed. Finally a short conclusion
is given in Sec. V.

\vskip 1pc \vskip 2pc \centerline{\large{\bf II. Model and Spin Wave
     Approximation}}
We consider a single central spin coupled to an
antiferromagnetic environment. It is assumed that the central spin
is a spin-$\frac{1}{2}$ atom and the antiferromagnetic
environment consist of atoms with spin $S$.
A global magnetic field is applied to
both the central spin and the antiferromagnetic environment. The
total Hamiltonian can be written as
\begin{eqnarray}
H&=&H_S+H_{SB}+H_B,
\end{eqnarray}
where $H_S$, $H_B$ are the Hamiltonians of the central spin and
the environment respectively, and $H_{SB}$ is the coupling term
\cite{17,18,33,45}. They can be written as
\begin{eqnarray}
H_S&=&-g\mu_BBS_0^z,\\
H_{SB}&=&-\frac{J_0}{\sqrt{N}}S_0^z \sum_{i}
(S_{a,i}^z+S_{b,i}^z),\\
H_B&=&J\sum_{i,\vec{\delta}}\mathbf{S}_{a,i}\cdot\mathbf{S}_{b,i+\vec{\delta}}
+J\sum_{j,\vec{\delta}}\mathbf{S}_{b,j}\cdot\mathbf{S}_{a,j+\vec{\delta}}\nonumber\\
&&-g\mu_B(B+B_A)\sum_{i} S_{a,i}^z-g\mu_B(B-B_A)\sum_{j}
S_{b,j}^z,
\end{eqnarray}
where $g$ is the gyromagnetic factor and $\mu_B$ is the Bohr
magneton. For simplicity, significant interaction (Eq. 3) between
the central spin and the environment is assumed to be of the Ising
type with $J_0$ being the coupling constant [20]. $J$ is the
exchange interaction and is positive for antiferromagnetic
environment. The effects of the next nearest-neighbor interactions
are neglected, although they may be important in some real
antiferromagnets.  We assume that the spin structure of the
environment may be divided into two interpenetrating sublattices
$a$ and $b$ with the property that all nearest neighbors of an
atom on $a$ lie on $b$, and \textit{vice versa} \cite{45}.
$\mathbf{S}_{a,i}$ ($\mathbf{S}_{b,j}$) represents the spin
operator of the $i$th ($j$th) atom on sublattice $a$ ($b$). Each
sublattice contains $N$ atoms. The indices $i$ and $j$ label the
$N$ atoms, whereas the vectors $\vec{\delta}$ connect atom $i$ or
$j$ with its nearest neighbors. $B$ represents a uniform external
magnetic field applied in the $z$ direction. The anisotropy field
$B_A$ is assumed to be positive, which approximates the effect of
the crystal anisotropy energy, with the property of tending for
positive magnetic moment $\mu_B$ to align the spins on sublattice
$a$ in the positive $z$ direction and the spins on sublattice $b$
in the negative $z$ direction.  
Reference \cite{Rossini07} has analyzed the crossover from the case of
a single link 
of the system spin to a bath spin, to the case of the central spin model   
in which the system spin is uniformly coupled to all the spins of the
bath. This may serve to justify the coupling Hamiltonian of Eq.~(3).

We use the Holstein-Primakoff transformation,
\begin{eqnarray}
S_{a,i}^+&=&\sqrt{2S-a_i^+a_i}a_i, \hspace*{2mm}S_{a,i}^-=a_i^+\sqrt{2S-a_i^+a_i}, \hspace*{2mm}S_{a,i}^z=S-a_i^+a_i,\\
S_{b,j}^+&=&b_j^+\sqrt{2S-b_j^+b_j}, \hspace*{2mm}
S_{b,j}^-=\sqrt{2S-b_j^+b_j}b_j,
\hspace*{3mm}S_{b,j}^z=b_j^+b_j-S,
\end{eqnarray}
to map spin operators of the environment onto bosonic operators.
We will consider the situation that the environment is
in the low-temperature and low-excitation limit such that
the spin operators in Eqs. (5) and (6) can be approximated as
$S_{a,i}^+\approx\sqrt{2S}a_i$, and
$S_{b,j}^+\approx b_j^+\sqrt{2S}$.
This can be justified as in this limit,
the number of excitation is small, and the
thermal average $<a_i^+a_i>$ and $<b_i^+b_i>$ is expected to be of
the order $O(1/N)$ and can be safely neglected with respected to
$2S$ when $N$ is very large.
The Hamiltonians $H_{SB}$ and $H_B$ can then be written in the
spin-wave approximation as \cite{43,44,45}
\begin{eqnarray}
H_{SB}&=&-\frac{J_0}{\sqrt{N}}S_0^z \sum_{i}(b_i^+b_i-a_i^+a_i),\\
H_B&=&E_0+2MSJ\left(\sum_{i}
a_i^+a_i+\sum_{j}b_j^+b_j\right)+2MSJ\sum_{i,\vec{\delta}}(a_ib_{i+\vec{\delta}}+a_i^+b_{i+\vec{\delta}}^+)\nonumber\\
&&+g\mu_B(B+B_A)\sum_{i} a_i^+a_i-g\mu_B(B-B_A)\sum_{j}b_j^+b_j,
\end{eqnarray}
where $M$ is the number of nearest neighbors of an atom and
\begin{eqnarray}
E_0=-2NMS^2J-2NSg\mu_BB_A.
\end{eqnarray}
We note here that in obtaining Hamiltonian (8)
in line with the approximations of 
$S_{a,i}^+\approx\sqrt{2S}a_i$, and
$S_{b,j}^+\approx b_j^+\sqrt{2S}$ in the low excitation limit, 
we have neglected any term
containing products of four operators. 
The low excitations correspond to
low temperatures, $T\ll T_N$, where $T_N$ is the N\'{e}el
temperature \cite{44}.

Transforming Eqs. (7) and (8) to the momentum space, we have
\begin{eqnarray}
H_{SB}&=&-\frac{J_0}{\sqrt{N}}S_0^z \sum_{\mathbf{k}}(b_\mathbf{k}^+b_\mathbf{k}-a_\mathbf{k}^+a_\mathbf{k}),\\
H_B&=&E_0+(2MSJ+g\mu_BB_A+g\mu_BB)\sum_{\mathbf{k}}
a_\mathbf{k}^+a_\mathbf{k}\nonumber\\
&&+(2MSJ+g\mu_BB_A-g\mu_BB)\sum_{\mathbf{k}}b_\mathbf{k}^+b_\mathbf{k}\nonumber\\
&&+2MSJ\sum_{\mathbf{k}}\gamma_{\mathbf{k}}(a_\mathbf{k}^+b_{\mathbf{k}}^++a_\mathbf{k}b_{\mathbf{k}}),
\end{eqnarray}
where
$\gamma_{\mathbf{k}}=M^{-1}\sum_{\mathbf{\vec{\delta}}}e^{i\mathbf{k}\cdot\mathbf{\vec{\delta}}}$.
Then by using the Bogoliubov transformation,
\begin{eqnarray}
\alpha_\mathbf{k}=u_\mathbf{k}a_\mathbf{k}-v_\mathbf{k}b_\mathbf{k}^+,\\
 \beta_\mathbf{k}=u_\mathbf{k}b_\mathbf{k}-v_\mathbf{k}a_\mathbf{k}^+,
\end{eqnarray}
where $u_\mathbf{k}^2=(1+\Delta)/2$,
$v_\mathbf{k}^2=-(1-\Delta)/2$, and
$\Delta=1/\sqrt{1-\gamma_{\mathbf{k}}^2}$, the Hamiltonians
$H_{SB}$ and $H_B$ can be diagonalized as ($\hbar=1$)
\begin{eqnarray}
H_{SB}&=&-\frac{J_0}{\sqrt{N}}S_0^z \sum_{\mathbf{k}}
(\beta_\mathbf{k}^+\beta_\mathbf{k}-\alpha_\mathbf{k}^+\alpha_\mathbf{k}),\\
H_B&=&E_0'+\sum_{\mathbf{k}}\omega_\mathbf{k}^{(+)}\left(\alpha_\mathbf{k}^+\alpha_\mathbf{k}+\frac{1}{2}\right)+\sum_{\mathbf{k}}\omega_\mathbf{k}^{(-)}\left(\beta_\mathbf{k}^+\beta_\mathbf{k}++\frac{1}{2}\right),
\end{eqnarray}
where $\alpha_\mathbf{k}^+$ ($\alpha_\mathbf{k}$) and
$\beta_\mathbf{k}^+$ ($\beta_\mathbf{k}$) are the creation
(annihilation) operators of the two different magnons with
wavevector ${\mathbf{k}}$ and frequency
$\omega_{\mathbf{k}}^{(+)}$ ($\omega_{\mathbf{k}}^{(-)}$)
respectively. We note here that the coupling between different
magnon modes under the spin-wave approximation is neglected. This
is a consequence of neglecting terms
containing products of four operators in Eq.~(8) in the
low temperature limit ($T\ll T_N$) of the spin-wave approximation, where
the number of excitations of the antiferromagnetic environment is
sufficiently small.
For a cubic crystal system in the small $k$ approximation,
\begin{eqnarray}
\omega_\mathbf{k}^{(\pm)}&=&2MSJ\sqrt{\left(1+\frac{g\mu_BB_A}{2MSJ}\right)^2+2\frac{k^2l^2}{M}-1}\pm
g\mu_BB,
\end{eqnarray}
where $l$ is the side length of cubic primitive cell of
sublattice. $E_0'$ is a new constant. Let $k=0$ and
$\omega_\mathbf{k}^{(-)}=0$, a critical magnetic field,
\begin{eqnarray}
B_c=\frac{2MSJ}{g\mu_B}\sqrt{\left(1+\frac{g\mu_BB_A}{2MSJ}\right)^2-1},
\end{eqnarray}
is obtained \cite{43,44,45}. When the external magnetic field exceeds
$B_c$, we have $\omega_\mathbf{k}^{(-)}<0$. This indicates that
this branch of magnon is no longer stable due to the externally
applied magnetic field. As a result, the antiferromagnetic
polarization flips perpendicular to the field, i.e., the magnetic
field induced spin-flop transition happens. The spin-flop
transition demonstrates a significant change of the spin
configuration in the antiferromagnetic environment. This
phenomenon has been observed and investigated for many different
materials \cite{Yunoki02,47,48}. 
The spin wave theory is known to describe well
the low-excitation and low-temperature properties of
antiferromagnetic materials. Despite this low-excitation
approximation, the spin wave theory also describes well the
physics for $B<B_c$ and the value of the critical magnetic field
of the spin-flop transition in antiferromagnetic materials
\cite{43,44,45}. We will thus use the spin wave theory to discuss the
decoherence time of the central spin under the influence of the
antiferromagnetic environment when the external magnetic field is
tuned to approach $B_c$ from below (i.e. $B\rightarrow B_c$ for
$B<B_c$). It will be shown that an analytic expression for the
decoherence time can be obtained.

\vskip 1pc \vskip 2pc \centerline{\large{\bf III. Decoherence Time
     Calculation}}

In this section, we calculate the time evolution of the off-diagonal
elements of the reduced density matrix for
the central spin.
We assume that the initial density matrix of the total systems is
separable, i.e., $\rho(0)=\rho_S(0)\otimes\rho_B$. The initial
state of the central spin is described by $\rho_S(0)$. The density
matrix of the environment satisfies a thermal distribution, that
is $\rho_B=e^{-H_B/T}/Z$, where $Z$ is the partition function and
the Boltzmann constant has been set to one. We are interested in
the dynamics of the off-diagonal elements of the reduced density
matrix which are responsible for the coherence of the system. This is
equivalent to calculating the time evolution of the spin-flip
operator $S_0^-=|0\rangle\langle1|$, where $|0\rangle$ and $|1\rangle$ are
respectively the lower and upper eigenstates of $S_0^z$.
By tracing out the environmental degrees of freedom, the time
evolution of $S_0^-=|0\rangle\langle1|$ can be written as:
\begin{eqnarray}
S_0^-(t)&=&\textrm{tr}_B\left\{e^{-iHt}\left[S_0^-(0)\otimes\rho_B
\right]e^{iHt}\right\}\nonumber\\
&=&\frac{1}{Z}\textrm{tr}_B
\left\{e^{-iHt}\left(|0\rangle\langle1|\otimes
e^{-H_B/T} \right)e^{iHt}\right\}\nonumber\\
 &=&|0\rangle\langle1|\frac{1}{Z}\textrm{tr}_B \left[e^{-ig\mu_Bt-i\frac{J_0}{\sqrt{N}} \sum_{\mathbf{k}}
(\beta_\mathbf{k}^+\beta_\mathbf{k}-\alpha_\mathbf{k}^+\alpha_\mathbf{k})t}e^{-H_B/T}
\right],
\end{eqnarray}
where the subscript $B$ of $\textrm{tr}_B$ indicates that the bath
degrees of freedom are traced over. 
We describe briefly how Eq.~(18) is obtained. 
Since the Hamiltonian $H_B$ commutes with the total Hamiltonian $H$, we
may move the last term in the second line of Eq.~(18) 
to the place next to the operator $|0\rangle\langle1|$ and evaluate
the resultant expression $e^{-iHt}|0\rangle\langle1|e^{iHt}$ first
using Eqs.~(1), (2), (14) and (15).
After this straightforward evaluation, the operator
$|0\rangle\langle1|$ can then be put outside of the trace
$\textrm{tr}_B$, as shown in the last line of 
Eq.~(18). This is because $\textrm{tr}_B$ does not
influence the degrees of freedom of the central spin. 
The partition function can be evaluated as follows.
\begin{eqnarray}
Z&=&\textrm{tr}_B\left(e^{-H_B/T}\right)\nonumber\\
&=&\textrm{tr}_B\left[e^{-E_0'-\sum_{\mathbf{k}}\omega_\mathbf{k}^{(+)}\left(\alpha_\mathbf{k}^+\alpha_\mathbf{k}+\frac{1}{2}\right)/T-\sum_{\mathbf{k}}\omega_\mathbf{k}^{(-)}\left(\beta_\mathbf{k}^+\beta_\mathbf{k}+\frac{1}{2}\right)/T}\right]\nonumber\\
&=&e^{-E_0'}
\prod_{\mathbf{k}}e^{-\omega_\mathbf{k}^{(+)}/2T}\prod_{\mathbf{k}}e^{-\omega_\mathbf{k}^{(-)}/2T}
\prod_{\mathbf{k}}\frac{1}{1-e^{-\omega_\mathbf{k}^{(+)}/T}}\prod_{\mathbf{k}}\frac{1}{1-e^{-\omega_\mathbf{k}^{(-)}/T}}.
\end{eqnarray}
Similarly, the last part of the numerator of Eq. (18) can be evaluated as
\begin{eqnarray}
\textrm{tr}_B \left[e^{-ig\mu_BBt-i\frac{J_0}{\sqrt{N}}
\sum_{\mathbf{k}}
(\beta_\mathbf{k}^+\beta_\mathbf{k}-\alpha_\mathbf{k}^+\alpha_\mathbf{k})t}e^{-H_B/T}
\right]
=e^{-ig\mu_BBt}e^{-E_0'}\prod_{\mathbf{k}}e^{-\omega_\mathbf{k}^{(+)}/2T}\nonumber\\
 \prod_{\mathbf{k}}e^{-\omega_\mathbf{k}^{(-)}/2T}\prod_{\mathbf{k}}\frac{1}{1-e^{i\frac{J_0}{\sqrt{N}}t}e^{-\omega_\mathbf{k}^{(+)}/T}}\prod_{\mathbf{k}}\frac{1}{1-e^{-i\frac{J_0}{\sqrt{N}}t}e^{-\omega_\mathbf{k}^{(-)}/T}}.
\end{eqnarray}
Defining decoherence factor $r(t)$ as
\begin{eqnarray}
S_0^-(t)=S_0^-(0)r(t),
\end{eqnarray}
we then obtain from Eqs.~(18)-(20),
\begin{eqnarray}
r(t)&=&\frac{y^+y^-}{y_0^+y_0^-}e^{-ig\mu_BBt},
\end{eqnarray}
where
\begin{eqnarray}
y_0^+&=&\prod_{\mathbf{k}}\frac{1}{1-e^{-\omega_\mathbf{k}^{(+)}/T}},\\
y_0^-&=&\prod_{\mathbf{k}}\frac{1}{1-e^{-\omega_\mathbf{k}^{(-)}/T}},\\
 y^+&=&\prod_{\mathbf{k}}\frac{1}{1-e^{i\frac{J_0}{\sqrt{N}}t}e^{-\omega_\mathbf{k}^{(+)}/T}},\\
 y^-&=&\prod_{\mathbf{k}}\frac{1}{1-e^{-i\frac{J_0}{\sqrt{N}}t}e^{-\omega_\mathbf{k}^{(-)}/T}}.
\end{eqnarray}
To further evaluate the decoherence factor in the thermodynamic
limit, we proceed as follows.
\begin{eqnarray}
\ln y_0^+&=&-\sum_\mathbf{k} \ln \left (1-e^{-\omega_\mathbf{k}^{(+)}/T} \right)\nonumber\\
&=&-\frac{V}{8\pi^3}\int \ln \left
(1-e^{-\omega_\mathbf{k}^{(+)}/T} \right)4\pi k^2dk,
\end{eqnarray}
where $V$ is the volume of the environment. At low temperature such
that $(\omega_\mathbf{k}^{(\pm)})_{max}>>T$, we may extend the
upper limit of the integration to infinity. With $x=kl$ and
$N=V/l^3$, we obtain
\begin{eqnarray}
\ln y_0^+&=&-\frac{N}{2\pi^2}\int_0^\infty \ln \left (
1-e^{-\omega_\mathbf{k}^{(+)}/T} \right)x^2dx.
\end{eqnarray}
In the same way, we have
\begin{eqnarray}
\ln y_0^-&=&-\frac{N}{2\pi^2}\int_0^\infty \ln \left (
1-e^{-\omega_\mathbf{k}^{(-)}/T} \right)x^2dx,\\
 \ln y^+&=&-\frac{N}{2\pi^2}\int_0^\infty \ln
\left ( 1-e^{i\theta}e^{-\omega_\mathbf{k}^{(+)}/T} \right)x^2dx,\\
 \ln y^-&=&-\frac{N}{2\pi^2}\int_0^\infty \ln
\left ( 1-e^{-i\theta}e^{-\omega_\mathbf{k}^{(-)}/T} \right)x^2dx,
\end{eqnarray}
where
\begin{eqnarray}
\theta=\frac{J_0}{\sqrt{N}}t.
\end{eqnarray}
Letting
\begin{eqnarray}
f^{\pm}(\theta)&=&\int_0^\infty \ln \left (
1-e^{\pm i\theta}e^{-\omega_\mathbf{k}^{(\pm)}/T} \right)x^2dx\nonumber\\
&&-\int_0^\infty \ln \left ( 1-e^{-\omega_\mathbf{k}^{(\pm)}/T}
\right)x^2dx,
\end{eqnarray}
we can rewrite the decoherence factor as
\begin{eqnarray}
r(t)=e^{-\left[\frac{f^+(\theta)+f^-(\theta)}{\theta^2}\right]\frac{J_0^2t^2}{2\pi^2}-ig\mu_BBt}.
\end{eqnarray}
Its absolute value is then
\begin{eqnarray}
|r(t)|=e^{-\textrm{Re}\left[\frac{f^+(\theta)+f^-(\theta)}{\theta^2}\right]\frac{J_0^2t^2}{2\pi^2}}.
\end{eqnarray}

In the thermodynamic limit, i.e., $N\rightarrow \infty$, we have,
from Eq. (32), $\theta\rightarrow 0$. It is then obvious from
Eq.~(33) that $f^\pm(\theta)\rightarrow 0$ as $\theta\rightarrow
0$. To find the relation between $\theta$ and
$\textrm{Re}f^{\pm}(\theta)$ in the thermodynamic limit, we
calculate
\begin{eqnarray}
\eta^{\pm}&=&\lim_{\theta\rightarrow0}\frac{\textrm{Re}f^{\pm}(\theta)}{\theta^2}\nonumber\\
&=&\textrm{Re}\lim_{\theta\rightarrow0}\frac{f^{\pm}(\theta)}{\theta^2}\nonumber\\
&=&\textrm{Re}\lim_{\theta\rightarrow0}\frac{\int_0^\infty \frac{\mp ie^{{\pm}i\theta}e^{-\omega_\mathbf{k}^{(\pm)}/T}}{1-e^{{\pm}i\theta}e^{-\omega_\mathbf{k}^{(\pm)}/T}}x^2dx}{2\theta}\nonumber\\
&=&\frac{1}{2}\int_0^\infty
\frac{e^{-\omega_\mathbf{k}^{(\pm)}/T}}{(1-e^{-\omega_\mathbf{k}^{(\pm)}/T})^2}x^2dx.
\label{eta_limit}
\end{eqnarray}
From the second to the third line of Eq.~(\ref{eta_limit}), we have
used the rule of $\lim_{x\to c}[f(x)/g(x)]=\lim_{x\to c}[f'(x)/g'(x)]$,
where the prime denotes the derivative.
The absolute value of the decoherence factor from Eq. (35) in the
thermodynamic limit
($N\rightarrow \infty$, i.e., $\theta\rightarrow0$) can then be expressed as
\begin{eqnarray}
|r(t)|=e^{-\frac{
J_0^2\left(\eta^++\eta^-\right)}{2\pi^2}t^2}=e^{-\frac{t^2}{\tau_0^2}},
\end{eqnarray}
where the decoherence time
\begin{eqnarray}
\tau_0=\frac{\sqrt{2}\pi}{J_0\sqrt{\left(\eta^++\eta^-\right)}}.
\end{eqnarray}
Equations (37) (38) and (36) are the central result in the present
paper. It indicates that the decoherence factor displays a
Gaussian decay with time. The factor $t^2$ in the exponent is different from
the Markovian approximation which usually shows a linear decay in
time in the exponent. The quantities $\eta^{\pm}$ in Eq.~(36) can in principle
be evaluated numerically. An analytic expression can, however, be
obtained [see Eq. (39)] as the external magnetic field approaches
the critical field value from the below, i.e., $B\to B_c$ and
$B<B_c$.

\vskip 2pc \centerline{\large{\bf IV. Results and Discussions}}

We discuss the numerical results and figures calculated
using Eqs.~(36)-(38), and present  
the analytical result when $B\to B_c$ from below in
this section.  For a typical antiferromagnetic anisotropy crystal,
one has $B_A\approx0.1\,\mathrm{Tesla}$,
$2MSJ/g\mu_B\approx100\,\mathrm{Tesla}$ \cite{43}. Our results obtained
using the spin wave approximation are valid only for not too large
temperatures, much smaller than the N\'{e}el temperature, $T_N$.
According to neutron diffraction studies, the N\'{e}el temperature
of antiferromagnet TbAuIn is $35\,\mathrm{K}$ \cite{49}. Some
antiferromagnets may have higher N\'{e}el temperatures.  So in the
following analysis, the environmental temperature is restricted
below $T/g\mu_B=2.5\,\mathrm{Tesla}$, i.e., $T\approx
3.4\,\mathrm{K}$. 

Figure 1 presents the decoherence time as a function of the
external magnetic field for different temperatures with a
crystal anisotropy field of $B_A=0.1\,\mathrm{Tesla}$. We can see that
the decoherence
time decreases with the increase of the external magnetic field.
Figure 2 is similar to Fig.~1, except $B_A=0.15\,\mathrm{Tesla}$.
Comparing Fig.~1 with Fig.~2, we find that the larger crystal anisotropy field
suppresses the decoherence of the central spin. This field-dependent
decoherence behavior may be inferred from the effective
Hamiltonian, Eqs.~(14)-(16).
From the interaction Hamiltonian, Eq.~(14), we see that the larger the
difference in the magnon excitation number between the magnon
$\beta_\mathbf{k}$ and magnon $\alpha_\mathbf{k}$, the stronger the
effect of the environment on the central spin. When the
external magnetic field is zero or absence,
the two magnons, from Eq.~(16), have the same
frequency for a given wavevector $\mathbf{k}$. At a given temperature,
the average
thermal excitation number may be the same for the two magnons, but the
fluctuation in the excitations for each individual magnon may not be
the same at the same time. As a result, tracing over the magnon states
still causes decoherence on the central spin.
If the external magnetic field is increased,
the magnon frequency  $\omega_\mathbf{k}^{(-)}$ decreases but
$\omega_\mathbf{k}^{(+)}$ increases. Consequently, the magnon mode
$\beta_\mathbf{k}$ is easier to be excited than the magnon mode
$\alpha_\mathbf{k}$ at a given anisotropy field and temperature.
This then results in a larger magnon excitation number difference
and fluctuation, and thus a stronger decoherence effect. So the
decoherence time decreases with the increase of the external
magnetic field. The effect of the anisotropy field on the
decoherence time could be understood in a similar reasoning. At a
fixed external magnetic field, both the magnon frequencies
$\omega_\mathbf{k}^{(-)}$ and $\omega_\mathbf{k}^{(+)}$ increase
with increasing anisotropy fields. Thus the average excitation
numbers reduce, so is the fluctuation in their difference. As a
consequence, the decoherence time increases with the increasing
anisotropy field.

\begin{figure}
\begin{center}
\includegraphics [width=9cm] {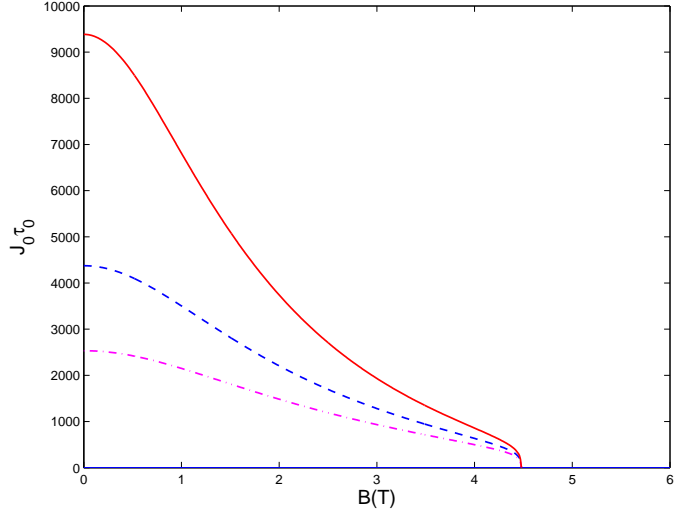}
\end{center}
\caption{The decoherence time as a function of the external
magnetic field for different temperatures:
$T/g\mu_B=0.8\,\mathrm{Tesla}$ (solid curve),
$T/g\mu_B=1.0\,\mathrm{Tesla}$ (dashed curve), and
$T/g\mu_B=1.2\,\mathrm{Tesla}$ (dot-dashed curve). Other
parameters are $M=6$, $MJ/g\mu_B=100\,\mathrm{Tesla}$,
$B_A=0.10\,\mathrm{Tesla}$.} \label{fig1}
\end{figure}

\begin{figure}
\begin{center}
\includegraphics  [width=9cm] {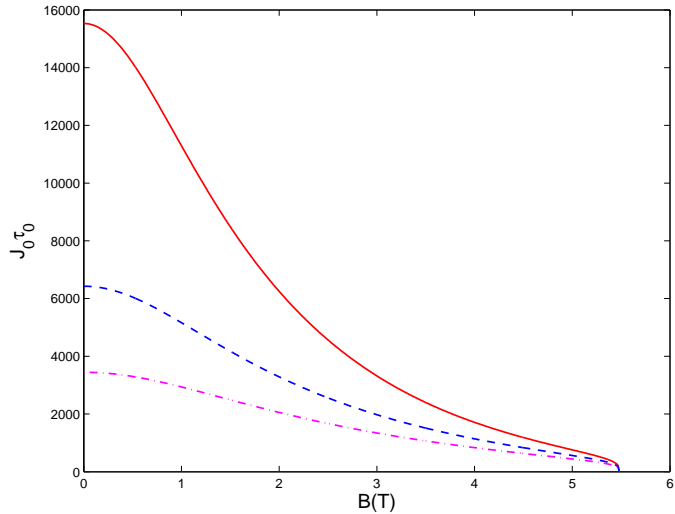}
\end{center}
\caption{Same as Fig. 1, except $B_A=0.15\,\mathrm{Tesla}$.}
\label{fig2}
\end{figure}

An alternative way to understand the field-dependent decoherence
time may be in terms of quantum correlations. There is a kind of
tradeoff between the external magnetic field and the anisotropy
field. The anisotropy field makes the antiferromagnets stable. On
the other hand, the external magnetic field tends to reduce the
antiferromagnetic order of the environment. Therefore the stronger
the external magnetic field, the smaller the antiferromagnetic
order. On the contrary, the larger the anisotropy field, the
stronger the correlation of the antiferromagnetic environment. If
the constituents (spins) of the environment maintain appreciable
correlations or entanglement between themselves, then there is a
restriction on the entanglement between the central spin and the
environment \cite{50,51}. As a consequence, this sets a restriction on the
amount that the central spin may decohere \cite{16,50,51,52}. Thus as far
as the decoherence of the central spin is concerned, the
anisotropy field has a similar effect to the exchange interaction
strength between the constituents (spins) of the antiferromagnetic
environment. Strong intra-environmental interaction results in a
strong antiferromagnetic correlation, thus an effective decoupling
of the central spin from the environment and a suppression of
decoherence \cite{50,51}. Therefore the decoherence time increases with
the increase of the anisotropy field but decreases with the
increase of the strength of the external magnetic field. In
summary, the results shown in Figs. 1 and 2 confirm that strong
correlations within the environment suppress the decoherence
effects \cite{16,50,51,52}.

In Ref.~\cite{Novais05}, a similar problem of
a spin-1/2 impurity coupled locally  
(in contrast to uniformly in the central spin model considered
here) in real space to an antiferromagnetic environment has
been studied in the context of quantum frustration of decoherence.
There, the Hamiltonian is reduced to the case of 
a localized impurity spin coupled
with two non-commuting spin component operators respectively to two
bosonic baths.  The quantum frustration is referred, in
Ref.~\cite{Novais05}, to the lack of a preferred basis 
for the impurity spin due to the competition of the non-commutativity
of the coupled spin operators to the two baths. 
As a result, the entanglement of the impurity spin with each one of
the baths is suppressed by the other, and the decoherence
phenomenon is thus frustrated. 
On the other hand, the effective Hamiltonian of Eqs.~(2), (14) and
(15) under the spin-wave approximation in our central spin model
indicates clearly that the $S_0^z$ eigenstates are the preferred basis
of the central spin. It also indicates that the influence of the two magnon
environments seems to partially cancel each other and lead to a situation
of less decoherence than a central spin coupled to a single bath. 
Though this may also be regarded as some kind of decoherence frustration,
its cause in this spirit seems different from the quantum frustration
of decoherence discussed in Ref.~\cite{Novais05}.

When the strength of the magnetic field is increased further
toward the critical point where the spin-flop transition occurs,
($B=B_c\approx4.5\,\mathrm{Tesla}$ for the parameters used in Fig.
1 and $B_c\approx5.5\,\mathrm{Tesla}$ in Fig. 2, the decoherence
time approaches zero quickly. One can obtain analytically from
Eqs.~(16), (36), and (38) for the decoherence time near the
critical field as
\begin{eqnarray}
\lim\limits_{B \rightarrow
B_c(B<B_c)}J_0\tau_0=\frac{4\sqrt{\pi}}{T}M^{\frac{3}{4}}J^{\frac{3}{2}}\frac{\left[g\mu_B(B_c-B)\right]^{\frac{1}{4}}}{\left(g\mu_BB_c\right)^{\frac{3}{4}}}.
\end{eqnarray}
We note that this result using spin wave theory is valid for
$B<B_c$, i.e., before the spin-flop transition. Equation (39)
indicates that when $B$ approaches $B_c$, the environment exerts a
great influence on the central spin so that the coherence is
destroyed thoroughly. This is a catastrophe for quantum computing
based on such spin systems at the critical field $B_c$. 
Our result in the thermodynamic limit is consistent with those in 
Refs.~\cite{Quan06,Cucchietti07,Rossini07,Zanardi07} for the case of 
finite spin number $N$.
There, the asymptotic value of the Loschmidt echo of a bath 
(or equivalently the decoherence factor of a central spin)
is dramatically reduced near the critical point and may serve as a
good witness of QPT in the case of finite $N$. 
In our case, the decoherence of the central spin is enhanced near the
critical point of the antiferromagnetic bath. The closer the bath to the
QPT (the spin-flop transition from below, $B<B_c$), 
the smaller the decoherence time of the central spin (see Figs.~1 and 2). 
The decoherence reaches its maximum when the values of the
external magnetic field is increased to the critical field point of
the spin-flop transition.   
The Gaussian decay behavior of the decoherence of a central spin
similar to Eq.~(37) has also been reported and 
discussed in Refs.~\cite{Quan06,Cucchietti07,Rossini07}.

We discuss the influence of the environmental
temperature on the decoherence behavior of the central spin below.
The decoherence behavior is sensitive to
the environmental temperature for weak external magnetic fields.
High temperatures cause strong decoherence (see Fig. 1). When the
external field is strong, the influence of the environmental
temperature becomes minor or weak. Finally, the decoherence time
is less sensitive to the change of the external magnetic field at
high temperatures. This can be seen from Fig. 1 that the low
temperature case (solid curve) has a sharper curve dependence than
the high temperature cases (dash and dot-dash curves).

\begin{figure}[htbp]
\begin{center}
\includegraphics  [width=9cm] {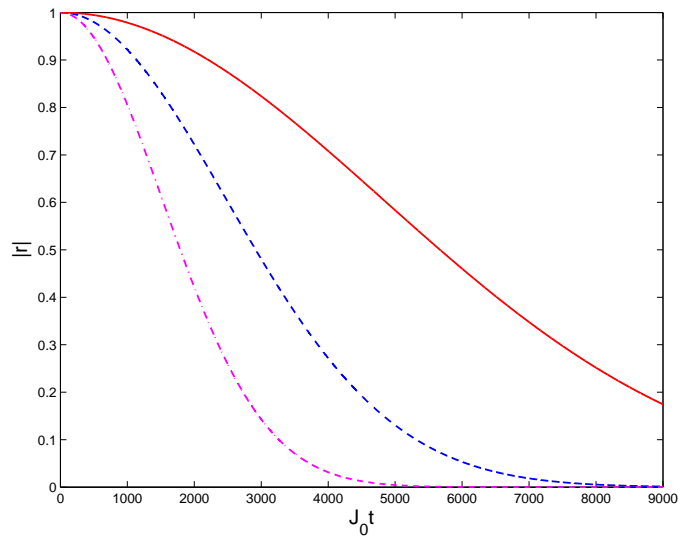}
\end{center}
\caption{Time evolution of the decoherence factor for different
temperatures: $T/g\mu_B=0.8\,\mathrm{Tesla}$ (solid curve),
$T/g\mu_B=1.0\,\mathrm{Tesla}$ (dashed curve), and
$T/g\mu_B=1.2\,\mathrm{Tesla}$ (dot-dashed curve). Other
parameters are $M=6$, $MJ/g\mu_B=100\,\mathrm{Tesla}$,
$B=1\,\mathrm{Tesla}$, $B_A=0.10\,\mathrm{Tesla}$.} \label{fig3}
\end{figure}

\begin{figure}[htbp]
\begin{center}
\includegraphics  [width=9cm] {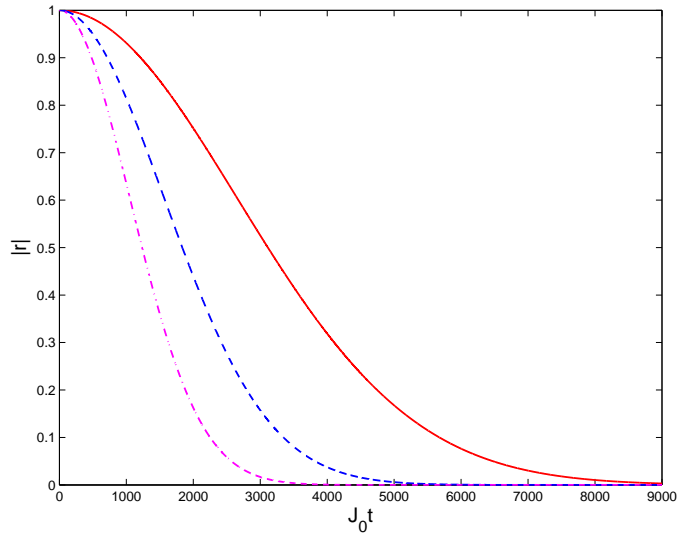}
\end{center}
\caption{Same as Fig. 3, except $B=2\,\mathrm{Tesla}$.}
\label{fig4}
\end{figure}

In Fig. 3, we plot the time evolutions of the decoherence factor
for different temperatures with the external magnetic field fixed
at $B=1\,\mathrm{Tesla}$. As expected, the decoherence factor,
under the influence of the environment, decreases with time. At a
higher temperature, the decay rate is larger as shown in the
figure. Figure 4 is similar to Fig. 3, except that the external
magnetic field is at $B=2\,\mathrm{Tesla}$. We see again that the
coherent behavior of the central spin is suppressed with the
increase of the external magnetic field.

\vskip 2pc \centerline{\large{\bf V. Conclusion}} We have studied
the decoherence of a central spin coupled to an antiferromagnetic
environment in the presence of an external magnetic field. The
results, obtained using the spin wave approximation in the
thermodynamic limit, show that the decoherence factor displays a
Gaussian decay with time. The decoherence time decreases, as
expected, with the increase of temperatures. Furthermore, the
external magnetic field promotes decoherence effects. The
decoherence reaches its highest point at the critical field of the
spin-flop transition, consistent with the QPT witness studies in 
Ref.~\cite{Quan06,Cucchietti07,Rossini07,Zanardi07}. 
In contrast, the strong anisotropy field
suppress decoherence of the antiferromagnetic environment on the
central spin. Therefore, in order to reduce the loss of coherence
of the the central spin, we could decrease the environmental
temperature, eliminate the external magnetic field, and choose the
antiferromagnetic surrounding or underlying antiferromagnetic
materials with a strong crystal anisotropy field.

\vskip 2pc \centerline{\large{\bf VI. Acknowledgments}} X.Z.Y. and
H.S.G. would like to acknowledge support from the National Science
Council, Taiwan, under Grants No. NSC95-2112-M-002-018 and No.
NSC95-2112-M-002-054, and support from the focus group program
of the National Center for Theoretical Sciences, Taiwan.
H.S.G. also acknowledges support from the
National Taiwan University under Grant No. 95R0034-02
and is grateful to the National Center for High-performance Computing,
Taiwan, for computer time and facilities. X.Z.Y. also acknowledges
support from the National Natural Science Foundation of China
under Grant No. 10647137.

\end{document}